# AI-Empowered Data Offloading in MEC-Enabled IoV Networks

Afonso Fontes[1], Igor de L. Ribeiro[2], *Khan Muhammad[3], Member, IEEE, Amir H. Gandomi[4], Senior Member, IEEE, Gregory Gay[1], Member, IEEE, Victor Hugo C. de Albuquerque[5,6], Senior Member, IEEE

[1]Chalmers and the University of Gothenburg, Gothenburg, Sweden
[2]University of Fortaleza, Fortaleza/CE, Brazil
[3]Department of Software, Sejong University, Seoul, Republic of Korea
[4]Faculty of Engineering & Information Technology, University of Technology Sydney, Australia
[5] Graduate Program on Teleinformatics Engineering, Federal University of Ceará, Fortaleza, Fortaleza/CE, Brazil
[6]Graduate Program on Electrical Engineering, Federal University of Ceará, Fortaleza/CE, Brazil

## Abstract

Advancements in smart vehicle design have enabled the creation of Internet of Vehicle (IoV) technologies that can utilize the information provided by various sensors and wireless communication to perform complex functionality. Many of these functionalities rely on high computational power and low latency. Mobile Edge Computing (MEC) technologies have been proposed as a way to meet these requirements, as their proximity and decentralization provide unique benefits for networks like real-time communication, higher throughput, and flexibility. Diverse challenges to the process of offloading data in a MEC-enabled IoV network have emerged, such as offloading reliability in highly mobile environments, security for users within the same network, and energy management to keep users from being disincentivized to participate in the network. This article surveys research studies that use AI as part of the data offloading process, categorized based on four main issues: reliability, security, energy management, and service seller profit. Afterward, this article discusses challenges and future perspectives for IoV technologies.

## 1. Introduction

With the advancement of new communication technologies, innovative utilization of various interconnected portable devices has emerged in smart cities, opening new possibilities for the automation of processes. Autonomous driving is one of the top trends in this context. Developments in AI techniques have allowed Intelligent Transportation Systems (ITS) to bring passengers and drivers a pleasant and safe environment [1]. These advancements lead to the possibility of autonomous driving that follows traffic rules and minimizes accidents caused by human errors. However, executing machine learning (ML)-based or image processing functionality on regular portable devices, typically with low computational power and limited battery autonomy, might not meet the quality of Service (QoS) requirements of such functionality.

Mobile Edge Computing (MEC), also known as fog computing [2], is a promising solution for IoV networks. Leveraging emerging wireless communication technologies, MEC is a widely distributed architecture that allows users to offload computation data from vehicles or any IoT device directly to nearby servers with "cloud-like" aptitudes. The closeness of MEC architectural components can provide high bandwidth, computational agility, and low latency, which can help significantly enhance safety and coordination between IoT and IoV devices, enabling numerous applications, especially in autonomous driving.

Although response time in a fog-like architecture is considerably faster than in cloud services, meeting QoS requirements is still challenging, mainly due to the high mobility of a vehicle. A vehicle may leave communication range before receiving a response from an offloaded task. This failure represents a worst-case scenario, once offloading data consumes battery/fuel from the vehicle and energy and computational power from the Roadside Unit (RSU) [3]. Therefore, how to design a framework that efficiently implements edge computing in extremely dynamic vehicular environments is still an open research challenge.

IoV and MEC have received increasing attention from researchers, and several studies examing various aspects of the field. For instance, Raza et al. [4] surveyed articles that address architecture, technical issues, and applications that utilize Vehicle Edge Computing (VEC) in MEC-enabled environments. Tong et al. [5] presented a survey of studies that uses AI and IoV for purposes that vary from simple sharing of traffic information to MEC-dependent tasks. A research niche on the usage of AI and Deep Learning (DL) for data offloading in MEC-enabled IoV networks is not investigated yet and needs exploration. Therefore, in this article, we examine how AI and DL based algorithms contribute to solving challenges related to offloading data in a MEC-enabled IoV network in recent research (since 2018).

This article is organized as follows. A background on MEC-enabled architectures is presented in Section 2. Section 3 presents the methodology adopted for selecting the studies examined in this review, followed by a description of the main issues related to offloading data in MEC-enabled networks in Section 4, highlighting essential components and limitations of each contribution. Section 5 examines open issues and research challenges in this field. Section 6 covers threats to validity of this work, and, finally, Section 7 concludes the article.

## 2. Background
In a MEC-enabled network, computing and storage services are provided at the edge of the network (close to the user), allowing services to be offloaded, computed, and responded to in a shorter time than traditional cloud computing. This fact alone makes edge computing appealing for IoV networks, as any significant delay affects critical tasks. In contrast, there are also tasks where time is less relevant than other factors like consistency or precision. For these tasks, QoS is the key to maximize the Quality of Experience (QoE).

Figure 1 depicts an example of offloading data in a MEC-enabled architecture based on [1, 2]. The bottom layer represents all devices connected in the network. In the middle, MEC devices serve as a link to the Cloud and the top layer represents Cloud computing itself. Notice that, in this structure, cloud technologies do not replace MEC or vice-versa. The Cloud can serve as a crucial component in several architectures (generally through delay tolerant information and tasks). The orange connections represent an IoT network, while the blue links show an IoV network. IoT networks connect remote devices, while IoV networks relate specifically to connected vehicles. With the emergence of smart and autonomous vehicles, there are a growing number of AI-dependent functionalities in IoV networks. Due to potential health risks (e.g., imminent collision), these functionalities are often critical and delay-sensitive. Figure 2 illustrates the possible communication channels in an IoV network, giving an idea of how volatile such a network can be: cars can communicate with each other via vehicle-to-vehicle (V2V) connections as well as with infrastructures along the road via vehicle-to-infrastructure (V2I) or even with a pedestrian's IoT device (V2P).

## 3. Methodology
Our interest in this literature review is to examine recent studies that use AI techniques to enhance data offloading management in IoV networks. We are interested in the reasons for and effects of such integration, which specific AI techniques have been adopted, and the potential impacts, risks, and insights of such AI integration.

Table 1 lists the research questions we aim to answer and briefly describes our motivation and objective in addressing each question. Questions 1, 2, and 3 allow us to understand how and why AI techniques have enhanced data offloading. RQ2 covers the authors' primary motivations. In contrast, RQ3 and RQ4 are technical questions, covering the role AI techniques play in data offloading and which techniques have been adopted. RQ5 covers open challenges in this field. Finally, RQ6 examines the benefits and limitations of adopting AI in offloading tasks (if the authors provide such information).

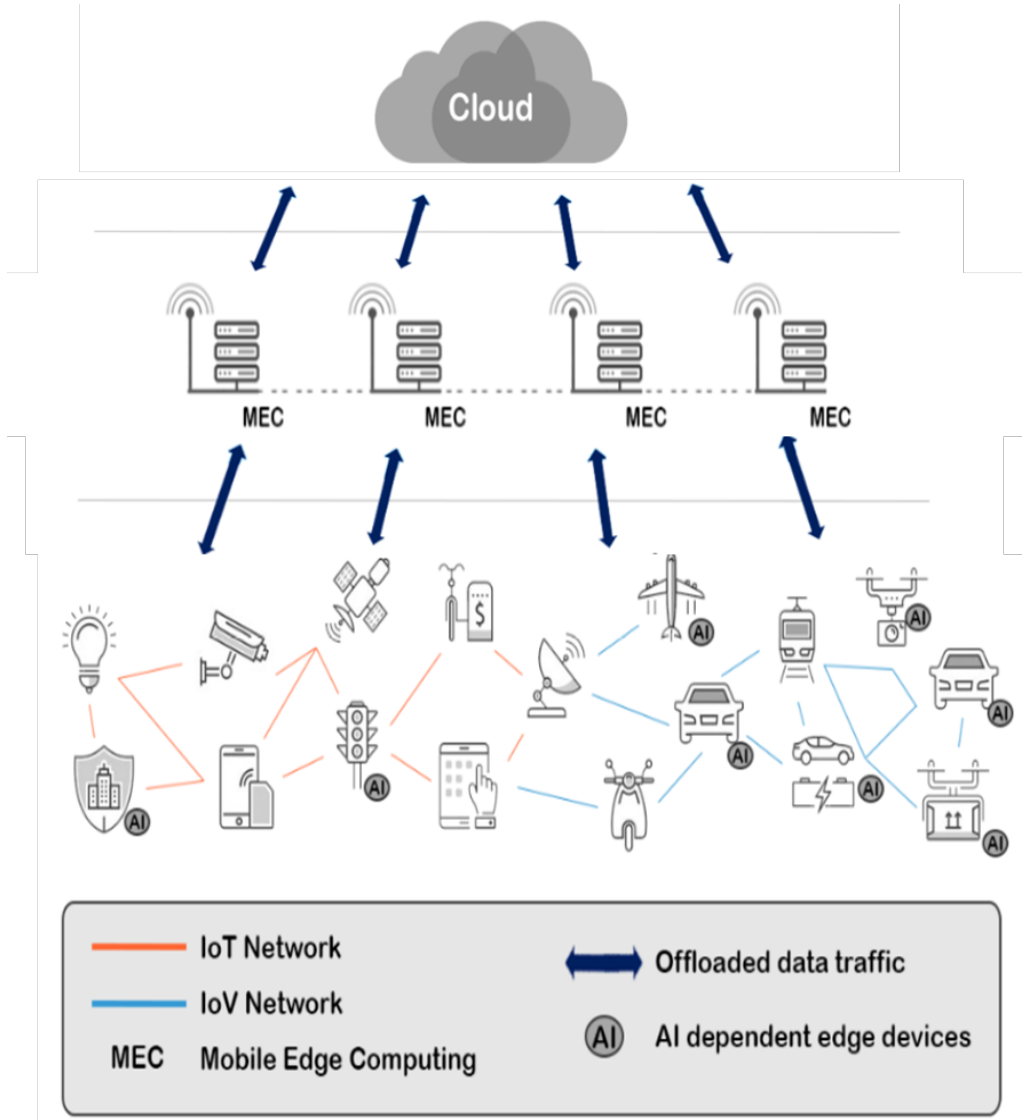

**Figure 1.** An example of a MEC-enabled network.

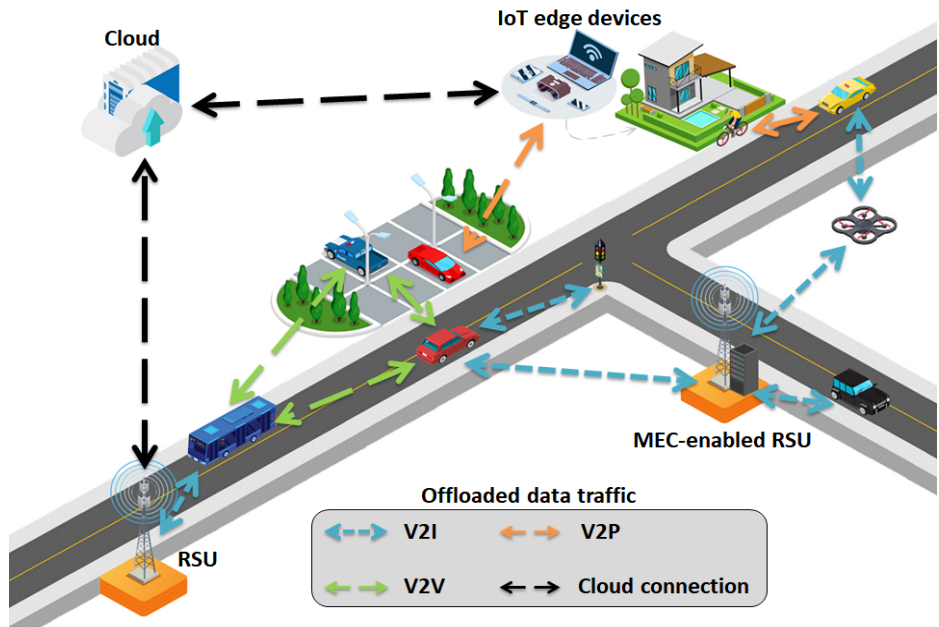

**Figure 2.** An overview of the MEC-enabled architecture in the field and representation of various communication channels used between devices, infrastructures, and other vehicles.

**Table 1.** List of Research Questions and their respective motivation/objective.

| ID | Research Question | Motivation |
|---|---|---|
| **RQ1** | What are the main challenges addressed by AI-empowered MEC-enabled networks? | Categorize the challenges addressed by adopting AI techniques. |
| **RQ2** | Why are these challenges being addressed? | Understand the motivatation of study authors. |
| **RQ3** | How were AI techniques employed to address the challenge? | List the specific tasks enhanced by adopting AI techniques. |
| **RQ4** | Which AI techniques have been adopted to address these challenges? | List the AI techniques employed. |
| **RQ5** | What are the open challenges and future trends in the use of AI techniques in data offloading in MEC-enabled networks? | Highlight the limitations of the examined research and open challenges in the area. |
| **RQ6** | What are the advantages and drawbacks of adopting AI techniques in MEC-enabled networks? | Describe the benefits and limitations of adopting AI. |

To gather the examined studies in this article, we conducted a search in the IEEE Xplore database for the following keywords: *"Artificial intelligence" AND ("offloading" OR "intelligent IoV") AND "mobile edge computing"*. In order to focus on recent contributions, we filtered the results by year, eliminating all entries older than 2018. The remaining articles were then reviewed by a specialist, who selected 11 entries to be analyzed in this survey.

## 4. Results and Discussion

This section presents how researchers have applied AI algorithms to address challenges related to offloading data in a MEC-enabled IoV network. Table 2 lists the studies that we analyzed, divided into four targeted offloading challenges: Reliability, Security, Energy Management, and Monetization. We also list the specific goal of applying AI to help address this challenge and note limitations of each approach. The following subsections further discuss each article in Table 2.

---

**RQ1: What are the main challenges addressed by AI-empowered MEC-enabled networks?**
The studies examined to address problems related to Reliability, Security, Energy Management, and Monetization.

---

The authors of these studies motivate the need for AI-empowered data offloading. The authors in [6, 7, 8, 9, 12] state that increased reliability is crucial for improving technologies and enabling new applications and functionalities. He et al. [10] highlight that as the usability of IoV and IoT networks grow, more threats will inevitably appear. Therefore, keeping the network secure is important to enlarge and encourage its usage. Ning et al. [11, 12] emphasize that energy is a crucial resource that must be managed in a mobile environment. Finally, the authors in [13, 14, 15] emphasize the importance of taking everyone's self-interests into account, especially in collaborative networks, to more accurately propose offloading schedules based on a realistic environment. Even though the highlighted issues tackle different aspects of the offloading task, they all are equally relevant to this emerging research field. If any of these four aspects is ignored, the capabilities of IoV networks will be limited.

---

**RQ2: Why are these challenges being addressed?**
To improve technologies, enable new applications, encourage usage, manage resources, improve simulations, and profit.

---

**Table 2.** Studies analyzed in this review, their targeted challenge, the algorithm adopted, the specific task the AI technique is responsible for, and specific noteworthy characteristics (i.e., strengths, limitations, or other relevant additional information).

| # | Targeted Challenge | Technique Used | AI Objective | Noteworthy Characteristics |
|---|---|---|---|---|
| [6] | Reliability | Deep Q-learning | Optimize the MEC server selection and minimize transmission faults by inserting redundancy protocols. | - Performance worsens as traffic density increases. Too much redundancy worsens offloading reliability. |
| [7] | Reliability | VE-MAN (Authorial) | Increase availability of the network and minimize latency time. | - Assumes every vehicle meets minimum hardware requirements to contribute to the network.<br>- Outperforms centralized offloading only if there are multiple vehicles in the network. |
| [9] | Reliability, Energy Management | Reinforcement Learning | Manage the use of UAVs as AP or computation nodes, positioning them according to the network's needs (Schedule and Path planning) | - The evaluation compares no UAVs, a UAV with a direct trajectory to a fixed point, and the planned trajectory given by the algorithm. |
| [8] | Reliability | Deep Reinforcement Learning (DRL) | Provide an optimal fluid/volatile policy for offloading directly from the field. | - Compared only with a greedy algorithm in an environment designed by the authors. |
| [12] | Reliability | FORT (Authorial) | Manage collaborative and distributed processing schedules using a simulated environment. | - Does not consider the node's energy consumption.<br>- Compared only with another proposed algorithm. |
| [10] | Security | PDS-learning | Outputs a protocol and a scheme for secure offloading. | - Increases energy consumption and offloading time. |
| [12] | Energy Management | Deep Reinforcement Learning (DRL) | Consider energy consumption as part of the problem. | - Compare multiple AI algorithms using real-world offloading data. |
| [11] | Energy Management | MEC-enabled Energy-Efficient Scheduling (MEES - Authorial) | Offloading schedule. Considers the direction and speed of a moving vehicle to decide where to compute. | - Minimizes energy consumption only in the MEC server.<br>- Assumes zero communication delay between RSUs.<br>- Compares an algorithm that can divide the offloaded task amongst other MEC servers with two others that cannot. |
| [14] | Monetization | Mobility Aware Double Deep Q-Network (Authorial) | Takes into consideration the mobility of vehicles, outputting fluid/volatile schemes that maximize QoE based on energy consumption and execution delay. | - Formulates a QoE equation.<br>- Compared with the greedy algorithm, DQN, Q-learning, and local computing. |
| [13] | Monetization | Game theory | Proposes an environment for comparing algorithms. Considers using a UAV as a communication bridge to enlarge the covered area for some time. | - Many assumptions were taken to create the game theoretic environment, which limits real-world applications. |
| [15] | Monetization | Markov Chain based algorithm | Integrates satellites to the IoV network. | - Builds robust simulation environment. |

## 4.1 Reliability

The high mobility of cars makes it challenging to apply MEC architectures to IoV networks. One of the worst scenarios in a MEC-enabled environment comes from a reliability fault where commands or results from a computationally heavy task do not arrive at the destination, resulting in an immense waste of processing power, time, and energy. The fact that reliability in data transfer processes is crucial was emphasized throughout every study analyzed. Several studies propose the use of AI to address reliability issues [6, 7, 8, 9, 12]. In this context, QoS requirements impose constraints on replying to tasks within the expected response time, depending on the information needed. For instance, in an autonomous vehicle, while a traffic update and the trajectory from location A to B could wait a few seconds, obstacle avoidance and movement prediction need to be computed as close to real-time as possible.

Several factors must be considered when choosing where to compute each task. An optimal solution must understand and manage resources, network access/limitations, user mobility, and the quantity of data to be offloaded. With that in mind, Zhang et al. [6] suggest the use of an in-depth Q-learning approach for designing offloading schemes that considers the selection of a target MEC server and determines the data transmission mode, highlighting the importance of a redundant offloading scheme in the case of vehicular data transmission failure. The authors also state that too much redundancy or too many fail-safe mechanisms cause extra traffic and might worsen reliability in a network as more devices connect. Wang et al. [2] and Qiao et al. [7] propose a vehicular network that increases performance in that scenario. They treat vehicles as edge computation resources and construct a collaborative distributed computing architecture. This approach is best suited for task offloading and data processing in urban environments, where there are many vehicles in proximity to each other. Collaborative task offloading and processing maximizes the use of IoV networks, resulting in a higher and more flexible coverage area. To optimize job scheduling and designate the best offloading transmission methods, both studies take into consideration the expected QoS for a specific task, the availability of neighboring vehicles, as well as MEC-enabled RSUs and a remote cloud server. However, the simulations assume that every vehicle meets a minimum hardware requirement. Qi et al. [8] propose the use of Deep Reinforcement Learning (DRL) to build an offloading strategy based on data collected on the go. If exposed to crowded urban conditions, the strategy used could eventually also result in a collaborative network. Nevertheless, in this approach, the topology strongly depends on the environment and what improves QoS.

To cope with the challenge that the high mobility of cars brings, Hu et al. [9] propose the use of unmanned aerial vehicles (UAVs) to enlarge the communication range of an RSU. The authors use a UAV as an access point, creating a communication bridge between the client and the service provider and, in some cases, also as a computation node. The study proposes a Deep Reinforcement Learning (DRL) algorithm to plan the position and the trajectory of the flying drone according to demand needs. As the battery is one of the main constraints for these devices, the strategy introduced in [9] is also discussed with regard to energy management in Section 4.3.

## 4.2 Security
The rapid development and increased supply of IoT and IoV devices impose unprecedented demands for communication and data processing in Cloud and fog-like structures. Several articles research privacy and security-aware protocols in communication from these devices to the Cloud. However, according to He et al. [10], this is not true when the transmission is local (fog-like).

He et al. [10] exploits the privacy vulnerability caused by wirelessly offloading tasks. In the studied scenario, the service provider is an agent that supplies computation and communication infrastructures to IoT devices, and each device can communicate with multiple service providers. The adversary in the system acts as a fake service provider in the network. An IoT device tends to connect to the nearest MEC server more frequently than others. With that knowledge, the adversary might use the location of the offloaded MEC server to infer the current activities of the user of that device. For instance, supposing the user connects to a MEC server installed within or near a bank, the adversary could infer their recent activities or, even using the same method, estimate the trajectory of a moving car by tracking its connection history. Both cases may lead to drastic consequences.

A possible solution to increase privacy and security in the offloading process is to add encryption to the communication protocols. The problem is that, as the complexity of messages increases, so grows the time to transfer and decode them, resulting in more energy consumption. He et al. use Post-Decision State (PDS)-learning, a Deep Neural Network (DNN) based algorithm, to produce privacy-aware offloading schemes that balance privacy and performance in a MEC-enabled network, resulting in a novel approach that allows the devices connected to cooperate, which, according to the authors, speeds up the learning curve. However, an external agent must manually set security thresholds for a given scenario in advance.

## 4.3 Energy Management
Autonomous vehicles leverage vast amounts of sensory information for smart navigation while posing intense computation demands. According to Ning et al. [11], each self-driving vehicle generates, on average, 30 TB of data per day. A potential solution to lower network delay and minimize response time is implementing decision-making ML and AI algorithms in RSUs equipped with MEC servers. However, these algorithms tend to consume large amounts of processing power and, therefore, energy, which may not always be available for edge devices. According to Ning et al. [12], energy shortage is becoming a crucial obstacle that limits the development of IoV.

To address the energy management challenge, Ning et al. [12] consider cloud servers, MEC-enabled RSUs and vehicles (parked and moving) as possible computation nodes. They seek to minimize the overall energy consumption for every node, especially if it is a moving vehicle, while satisfying the delay constraint for each specific task. Ning et al. [12] construct and train a DRL algorithm to output offloading schemes that solve this optimization problem based on the queuing theory. This approach considers the expected processing/response time and the type of the offloaded data before deciding where and how it should be processed. Ning et al.[11] focus on the energy management of RSUs. They propose a scheduling framework that, if needed, can distribute the task amongst other RSUs along the road in the direction of the moving vehicle, which ends up increasing reliability without enlarging the communication message with redundant protocols. According to the authors, the suggested framework shows low average energy consumption; however, for its tests and comparisons, [11] consider zero energy consumption and no delay while transferring data from one RSU to another.

Hu et al. and Asheralieva et al. propose using UAV [9, 13] to increase the coverage area of a centralized MEC-enabled network, which adds one more element to the energy management problem. Qi et al. [8] examine a MEC architecture that includes UAVs that may serve as a computing server or as a bridge for further offloading the computation tasks to an Access Point (AP). Because IoV applications and UAVs are usually power-limited, the major problem this approach encounters, is energy. The modeled system uses successive convex approximation (SCA) methods to minimize the energy consumption of the UAV and the User Equipment (UE) by managing bandwidth allocation and the UAV's current position. The algorithm proposed in [9] outputs an optimal trajectory for the flying drone that maximizes coverage and minimizes energy consumption. The authors demonstrate that the UAV's trajectory is greatly affected by the relative location of APs and UEs in the simulated environment and by offloaded task sizes. When managing computation-heavy latency-critical tasks, the benefits of the suggested algorithm gain relevance.

Several studies have presented a collaborative task offloading scheme where the computing is made by user vehicles instead of centralized computing offloading schemes. However, most studies do not consider expenses such as energy cost and battery depreciation the vehicles performing the edge computing task will inevitably have, especially when performing a large quantity of complex tasks. To address these costs, it is also essential to consider the possibility of a node selling unused processing power. Ning et al. [14] and Li Wang et al. [15] propose this scenario, adding the possibility for each node to profit from executing tasks or sharing their idle computational resources while also encouraging resource-rich agents to maximize income.

Following the same line of study, Asheralieva et al. [13] present a framework for computational offloading in a network formed by MEC servers installed at stationary base stations (BSs) and adds multiple UAVs agents acting as almost-stationary BSs. The location of a UAV could vary between simulations, but, unlike [9], the coverage area of the UAV is not moving. It remains stationary for one hour and then returns to the base for 10 minutes, simulating the time the battery would last and the time needed to recharge/replace the battery. The authors also take into consideration that the BSs can be owned by different service providers, which leads each simulated entity to favor the actions that maximize their provider's profit and cooperate with other BSs only when suitable. This approach results in a non-optimal and realistic simulation of what would happen in the real world.

Asheralieva et al. [13] model the environment and network as a two-level hierarchical game, supported by the game-theoretic (GT) and reinforcement learning (RL) techniques. The upper level defines the process of combination formation, and the lower level represents the set of self-interested behavior a BSs could have. An RL algorithm is responsible for learning how to manage every resource, especially energy, to maximize a service provider's payoff in each possible state the BSs could assume.

## 4.4 Monetization

Many studies present task offloading approaches in simulated environments intended to solve problems related to the device's requirements, network access, user mobility, and response time. However, in real-world applications, additional factors must be considered when computing data dependency tasks in a MEC-enabled network, like the monetary viability of the solution. Constrains like the initial investment needed, energy cost, and maintenance cost are amongst the most important factors to consider when discussing the viability of MEC. The authors of [13, 14, 15] propose a realistic scenarios by adding determinants like self-interested service providers and the possibility for users to receive income from sharing idle computational resources.

In the game-theoretic approach proposed by Asheralieva et al. [13], every computational node, including moving vehicles, acts as a player that competes for task offloading services to get higher utility, and therefore, profit. In this approach, each agent independently determines its offloading strategy based on self-interest and might neglect the interaction between multiple nodes. In terms of reliability and response time, the overall system performance might get worse when compared with purely cooperative systems. However, the scenario depicted may be more realistic than a purely cooperative case.

The algorithm proposed by Ning et al. [14] aims to produce offloading schemes that increase the average profit earned by service sellers from vehicles. The authors develop a two-step algorithm, which first outputs a matching scheme designed to schedule offloading requests that maximize the number of completed tasks from vehicles by allocating network resources. Then it uses a Double Deep Q-Network (DDQN) to manage resource allocation to enhance profit.

Li Wang et al. [15] propose an Incentive Mechanism that divides users into two categories: buyers and sellers. The buyers have computation-intensive applications waiting to be executed, and the sellers are nodes with idle computational power. This approach allows sellers to receive revenue by increasing the utilization of idle resources. The authors conclude that the proposed mechanism decreases average application completion time when compared to centralized MEC. Motivated by the fact that a seller gets rewarded, nodes willingly provide idle resources, making a collaborative approach more appealing.

## 5. Discussion

Modern vehicles embedded with sensors, wireless communication, and processing capabilities offer several benefits. However, such applications rely on high computational power and low latency. MEC offers a solution as its proximity and decentralization provide unique benefits for networks like real-time communication, higher throughput, and flexibility. However, diverse challenges in the process of offloading data in a MEC-enabled IoV network remain to be solved.

Table 2 lists recent studies on data offloading, published since 2018. Each of these studies offer essential contributions to the process of offloading data in an IoV network. However, they lack a standard simulation tool or dataset to test and compare in the same environment. In general, each team tend to narrow the tests to scenarios that favor their proposed approach. Although Asheralieva et al. [13] construct a game theory simulated environment that seems to be universal (and could potentially be used by all others), they make several assumptions that limit real-world application.

Many of the studies enhance an existing algorithm or framework. In most cases, the articles lack robust evaluation. The techniques proposed are often compared with the existing algorithms they are improving and other simplistic algorithms, and both tend to get outperformed in every test.

---

**RQ3: How were AI techniques employed to address the challenges?**
Optimizing offloading route and server selection; Generating centralized and distributed schemes; To enhance redundant and secure protocols; Latency management; Path planners for mobile nodes; Resource management and profit optimization.

---

> **RQ4: Which AI techniques have been adopted to address these issues?**
> Deep Q-Learning, RL (and the RL-based VE-MAN, FORT, MEES), DRL, Double Q-Network, Game Theory, and Markov Chain.

IoV and edge computing are trending topics that have been influenced directly by new technologies in several fields like communication, processing power, and battery lifetime. As these emerging technologies arrive, new research contributions tend to introduce different perspectives to address the offloading issues mentioned in this article.

1) *Universal Simulation Environment*: Current IoV-centered studies focus on achieving better computing performance through novel algorithms or frameworks. However, each study constructs a unique simulated environment, often making use of different assets/libraries, which difficult even more the comparison between frameworks and the assessment of how impacting the proposed systems are. Therefore, a universal simulation tool or environment would be of great value.

2) *Effects on Mobile Devices*: Current research involving MEC in IoT environments presume the use of mobile devices as added computational resources. However, the effects of this added computational load on mobile devices and how it affects the device's battery and lifespan may reduce the user's willingness to cooperate. Therefore, research in this direction is recommended to ensure the applicability of MEC networks.

> **RQ5: What are the open challenges and future trends in the use of AI techniques in MEC-enabled networks?**
> Techniques lack robust evaluation and a comparison tool/procedure; No contribution offered or used a publicly available database; The presented works are not reproducible by the readers; As the use of IoV networks increases, more demands will appear as well as more investment and space for new approaches, frameworks, and techniques.

In the analyzed studies, authors point out that the benefits of adopting AI into offloading tasks in MEC-enabled IoV networks outweigh the negative aspects of the adoption. AI-based algorithms tend to enable near-optimal management of the tasks on interest (e.g., generation of offloading schemes, server selection, etc.). However, this comes at the cost of higher processing power requirements and, therefore, more energy consumption. However, no authors conducted an in-depth experiment to measure the exact impact of incorporating.

> **RQ6: What are the advantages and drawbacks of adopting AI techniques in MEC-enabled networks?**
> Better management of the tasks is enabled. However, this comes at the cost of processing power and energy consumption.

## 6. Threats to Validity
Our search methodology used the IEEE Xplore database as the only source of articles. The keywords used in the search could include more terms, for instance: Neural Network (NN), Machine Learning (ML), and other variations. A comprehensive search without the publication year filter could be performed to (hopefully) increase the relevancy level of the articles analyzed. No snowballing technique was performed. The final list of papers to be analyzed was entirely based on a single specialist's opinion.

## 7. Conclusion
MEC delivers a high level of scalability, real-time data transfer, and mobility support, which are characteristics that suit the needs of an IoV network especially for tasks that require real-time decision-making. Although MEC enables complex vehicular applications, managing and keeping QoS, there are still numerous challenges related to data offloading. This literature review summarized recent studies (since 2018) that use AI algorithms to schedule data offloading and manage distributed computational resources to meet QoS requirements. These approaches are intended to maximize reliability, security, energy management, and profit for service sellers.

## Acknowledgments

This work was supported by the National Funding from the FCT - Fundação para a Ciência e a Tecnologia, through the UID/EEA/50008/2019 Project; and by Brazilian National Council for Research and Development (CNPq) via Grants No. 309335/2017-5; 304315/2017-6 and 430274/2018-1.

## Authors short bios

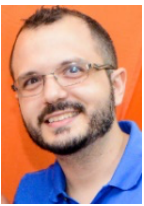

**Afonso Fontes** (afonso.fontes@chalmers.se) graduated as a Control and Automation Engineer with an emphasis in Robotics and has a Master of Science degree in Computer Science, in the line of study of artificial intelligence and software development. He is an experienced Lecturer and Researcher in the areas of Robotics, Autonomous Systems, and Digital Control. Afonso is currently a Ph.D. student at Chalmers and the University of Gothenburg, developing research in the use of Machine Learning techniques to solve Software Testing problems.

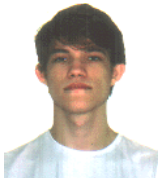

**Igor de L. Ribeiro** (firstbrhere@edu.unifor.br) is an undergraduate in Control and Automation Engineering at the University of Fortaleza (UNIFOR). His research interests include Swarm Technology and Artificial Neural Networks.

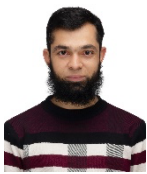

**Khan Muhammad [S'16, M'18]** (khan.muhammad@ieee.org) is an assistant professor at the Department of Software and Director of Visual Analytics for Knowledge Laboratory (VIS2KNOW Lab), Sejong University, South Korea. His research interests include video summarization, computer vision, big data analytics, IoT, 5G, intelligent transportation, and video surveillance. He has authored over 180 papers in peer-reviewed international journals, such as IEEE COMMAG, NETWORK, TII, TIE, IoJT, TNNLS and TSMC-Systems, and is a reviewer of over 100

SCI/SCIE journals, including IEEE COMMAG, WCOMM, NETWORK, IoTJ, TIP, TII, TCYB, Access and ACM TOMM. He is an Associate Editor/EB of several journals.

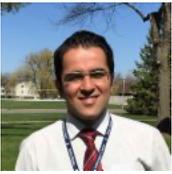

**Amir H. Gandomi** is a Professor of Data Science and an ARC DECRA Fellow at the Faculty of Engineering & Information Technology, University of Technology Sydney. Prior to joining UTS, Prof. Gandomi was an Assistant Professor at Stevens Institute of Technology, USA and a distinguished research fellow in BEACON center, Michigan State University, USA. Prof. Gandomi has published over two hundred journal papers and seven books which collectively have been cited 22,000+ times (H-index = 69). He has been named one of the most influential scientific minds and Highly Cited Researcher (top 1% publications and 0.1% researchers) for four consecutive years, 2017 to 2020. He also ranked 18th in GP bibliography among more than 12,000 researchers. He has served as associate editor, editor, and guest editor in several prestigious journals like AE of IEEE TBD and IEEE IoTJ. Prof Gandomi is active in delivering keynotes and invited talks. His research interests are global optimization and (big) data analytics using machine learning and evolutionary computations.

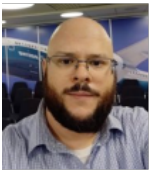

**Gregory Gay** (ggay@chalmers.se) is an Assistant Professor in the Interaction Design and Software Engineering division at Chalmers and the University of Gothenburg. Prior to this role, he was an Assistant Professor at the University of South Carolina. His research interests include automated testing and analysis and search-based software engineering. He has published over 40 publications in major SE venues, and recently received the 2009-2019 Most Influential Paper award at the IEEE International Conference on Software Maintenance and Evolution. Dr. Gay has served on the organizing and steering committees of ICSME, ICST, ASE, SSBSE, and others.

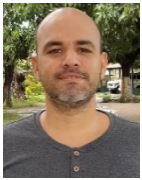

**Victor Hugo C. de Albuquerque [M'17, SM'19]** (victor.albuquerque@unifor.br) completed Ph.D. in Mechanical Engineering from the Federal University of Paraiba (UFPB, 2010), M.Sc. in Teleinformatics Engineering from the Federal University of Ceará (UFC, 2007). He is currently Assistant VI Professor of the Graduate Program in Applied Informatics at the University of Fortaleza (UNIFOR). He has authored or co-authored over 160 papers in refereed international journals, conferences, four book chapters, and four patents. He is an Editorial Board Member and Lead Guest Editor of several high-reputed journals, and TPC for many international conferences.